\documentclass[epj]{svjour}
\usepackage{graphicx}
\usepackage{subfigure}
\usepackage{color}
\usepackage{amsmath}

\def\mathbi#1{\textbf{\em #1}}

\begin{document}
\title{The prospects of sympathetic cooling of NH molecules \\ with Li atoms}
\author{Alisdair O. G. Wallis \and Edward J. J. Longdon
\and Piotr S. \.Zuchowski \and Jeremy M. Hutson}
\institute{Department of Chemistry, Durham University, South
Road, Durham, DH1~3LE, United Kingdom}

\date{\today}

\abstract{We calculate the quartet potential energy surface for
Li+NH and use it to calculate elastic and spin-relaxation cross
sections for collisions in magnetically trappable
spin-stretched states. The potential is strongly anisotropic
but spin-relaxation collisions are still suppressed by
centrifugal barriers when both species are in spin-stretched
states. In the ultracold regime, both the elastic and inelastic
cross sections fluctuate dramatically as the potential is
varied because of Feshbach resonances. The potential-dependence
is considerably reduced at higher energies. The major effect of
using an unconverged basis set in the scattering calculations
is to shift the resonances without changing their general
behaviour. We have calculated the ratio of elastic and
spin-relaxation cross sections, as a function of collision
energy and magnetic field, for a variety of potential energy
surfaces. Most of the surfaces produce ratios that are
favorable for sympathetic cooling, at temperatures below about
20~mK.}

\PACS{{34.50.Cx}{}}

\maketitle

%\fbox{\parbox{8cm}{{\bf Notes for copy editor:} \hfil \break We
%have been extremely careful with the italicisation of
%subscripts/superscipts and use Roman type for sub/superscripts
%that are abbreviations and italic type for those that are
%mathematical indices. Please do not change this unnecessarily.
%\hfil\break Upper-case Greek letters used in molecular term
%symbols should really not be slanted as produced by the svjour
%style.}}

\section{Introduction}

Since 1995, when the first gaseous Bose-Einstein condensates
(BECs) were created \cite{ANDERSON:1995,DAVIS:1995}, the
ability to cool, trap, and control atoms at ultracold
(sub-millikelvin) temperatures has increased dramatically. More
recently, attention has turned to cold and ultracold molecules
\cite{Doyle:2004,Krems:IRPC:2005,Friedrich:wacmsh:2009}. With
their more complex energy structure, ultracold molecules open
up many new possibilities in fields such as high-precision
measurement \cite{Hudson:2002,bethlem:FD142:2009}, quantum
computing \cite{DeMille:2002}, and ultracold chemistry
\cite{bell:2009}. However, this extra complexity comes at a
price, with the extra structure making molecules more difficult
to cool.

Atoms are usually cooled to microkelvin temperatures by laser
Doppler cooling \cite{Phillips:1998}. However laser cooling is
much less effective for molecules than for atoms. For this
reason, many other methods for cooling molecules have been
developed. These can be categorized into two distinct types,
direct and indirect methods
\cite{Bethlem:IRPC:2003,Hutson:IRPC:2006}. Indirect methods
involve the formation of ultracold molecules from a sample of
precooled ultracold atoms, and deeply bound and absolute
ground-state alkali-metal dimers have recently been formed, by
magnetoassociation (Feshbach resonance tuning) followed by
laser state transfer with Stimulated Raman Adiabatic Passage
(STIRAP)
\cite{Danzl:v73:2008,Lang:ground:2008,Ni:KRb:2008,Danzl:2010}.
Indirect methods are limited by the range of ultracold atoms
available, mainly to molecules formed from alkali-metal atoms.
Direct methods, such as buffer-gas cooling
\cite{Weinstein:CaH:1998} and Stark deceleration
\cite{Bethlem:IRPC:2003}, involve directly cooling a sample of
warm molecules down towards ultracold temperatures. NH
molecules have been buffer-gas cooled and magnetically trapped
in their X$^3\Sigma^-$ state, both alone
\cite{Egorov:2004,Campbell:2007,Tsikata:2010} and in
combination with N atoms \cite{Hummon:2008}, and
Stark-decelerated and electrostatically trapped in their
metastable a$^1\Delta$ state \cite{Hoekstra:2007}. Proposals
exist to transfer the metastable a$^1\Delta$ molecules into the
X$^3\Sigma^-$ ground state, either with an intense laser field
\cite{vandeMeerakker:2001} or by collisions with Rb atoms
\cite{Haxton:2009}. However, the lowest temperatures currently
obtainable by direct methods are tens to hundreds of
millikelvin, so that a secondary cooling stage is required to
reach ultracold temperatures.

One possible secondary cooling method is sympathetic cooling,
in which precooled cold molecules are placed in contact with an
ultracold atomic coolant and allowed to thermalize. However,
for sympathetic cooling to be effective, the rate of elastic
(cooling) collisions must greatly exceed the rate of inelastic
(trap-loss and heating) collisions \cite{deCarvalho:1999}. As a
rule of thumb, the ratio of elastic to inelastic cross sections
must exceed 100. Molecules trapped in magnetic and
electrostatic traps are held in low-field-seeking states
(states in which the energy increases with increasing field).
Collisional spin-relaxation processes from low-field to
untrapped high-field-seeking states will therefore limit the
efficiency of sympathetic cooling, and in many of the systems
theoretically studied the rate of collisional trap loss will be
too rapid for sympathetic cooling to work. Sympathetic cooling
has yet to be demonstrated for neutral molecular systems.
However it has been used to cool trapped ions
\cite{Larson:1986,Ostendorf:2006}, to produce overlapping
Bose-Einstein condensates \cite{Myatt:1997}, and to cool
fermionic \cite{Truscott:2001} and bosonic \cite{Modugno:2001}
atoms to quantum degeneracy.

The first to consider the sympathetic cooling of molecules with
alkali-metal atoms were Sold\'{a}n and Hutson
\cite{Soldan:2004}, who examined the interaction between
rubidium atoms and NH(X$^3\Sigma^-$). They found that the
dispersion-bound states of RbNH that correspond to the
Rb($^2S$)+NH(X$^3\Sigma^-$) dissociation threshold are crossed
by deeper ion-pair states, which correspond to the
Rb$^+$($^1S$) + NH$^-$($^2\Pi$) dissociation threshold. These
ion-pair states introduce mechanisms for inelastic collisions
and three-body recombination that may hamper sympathetic
cooling. Lara {\em et al.} subsequently investigated the
Rb($^2S$)+OH($^2\Pi_{3/2}$) system
\cite{Lara:PRL:2006,Lara:PRA:2007} and also found ion-pair
states that crossed the RbOH covalent states. They also carried
out full quantum calculations of field-free low-energy
collisions on the coupled potential energy surfaces and found
fast spin relaxation collisions that would prevent sympathetic
cooling for most states. More recently, detailed potential
energy surfaces and field-free scattering calculations for
Rb+NH and Cs+NH were performed by Tacconi {\em et al.}\
\cite{Tacconi:a:2007,Tacconi:b:2007,Tacconi:c:2007}, who found
large rotational inelasticities.

Sympathetic cooling of NH$_3$ and ND$_3$ has also been
considered. \.Zuchowski and Hutson \cite{Zuchowski:NH3:2008}
surveyed the interaction potentials of NH$_3$ with alkali-metal
and alkaline-earth atoms. They found that all the systems
studied exhibited large anisotropies that would produce strong
rotation-tunnelling inelasticity, so that sympathetic cooling
is unlikely to be successful for NH$_3$ in low-field-seeking
states. \.Zuchowski and Hutson \cite{Zuchowski:NH3:2009} went
on to examine the collisions of NH$_3$ and ND$_3$ with Rb,
showing that there is a good prospect of sympathetic cooling of
ND$_3$ in high-field-seeking states, even in collision with
magnetically trapped atoms.

Lara {\em et al.}\ \cite{Lara:PRA:2007} identified several desirable
features for sympathetic cooling of molecules that are not in their
absolute ground state. First, it is desirable to use light atomic
cooling partners, which produce high centrifugal barriers that can
suppress inelastic loss in channels with a small kinetic energy
release. Secondly, it is desirable to use molecules where the electron
spin is weakly coupled to the intermolecular axis, such as molecules
with Hund's case (b) coupling. Thirdly, it is desirable to use
atom-molecule systems for which the interaction potentials are only
weakly anisotropic. They suggested that closed-shell atoms such as the
alkaline earths might make good collision partners for magnetically
trapped molecules.

In 2009 Sold\'{a}n, \.Zuchowski, and Hutson
\cite{Soldan:MgNH:2009} followed up these suggestions: they
chose the case (b) molecule NH(X$^3\Sigma^-$) as a test case
and investigated its interactions with both alkali-metal and
alkaline-earth atoms. All the alkali-metal and the heavier
alkaline-earth systems (Ca-NH and Sr-NH) were found to have
strongly anisotropic interaction potentials, but Be-NH and
Mg-NH were found to have weak anisotropy. All the systems have
ion-pair states, but for Be and Mg the ion-pair states cross
the dispersion-bound state high up on the repulsive wall, so
that the ion-pair states will be energetically inaccessible in
low-temperature collisions. Wallis and Hutson
\cite{Wallis:PRL:MgNH:2009} then performed detailed quantum
scattering calculations on Mg+NH(X$^3\Sigma^-$) and showed that
the ratio of elastic to inelastic cross sections is large
enough to allow sympathetic cooling over a wide range of
collision energies and magnetic fields. Sympathetic cooling of
NH with Mg has a good prospect of success if the molecules can
be precooled to 10 mK and good trap overlap can be achieved.

Part of the success of Mg as a coolant comes from the fact that
its ground state has zero electron spin, so that only spin
relaxation and not spin exchange is possible in collisions with
NH. However, the lack of electron spin also prevents magnetic
trapping. In addition, Mg is quite hard to laser-cool, though
Mehlst\"aubler {\em et al.}\ \cite{Mehlstaubler:2008} have been
successful in achieving sub-Doppler temperatures. An
alternative way to prevent spin exchange is to use an
alkali-metal atom as the collision partner, while ensuring that
all collisions are between atoms and molecules in
spin-stretched states, with the maximum possible values of both
electron and nuclear spin projection quantum numbers. The use
of spin-stretched states also prevents strong interactions
involving ion-pair states. It is important to know whether this
approach would result in low enough inelastic collision rates,
despite the much larger anisotropies in alkali-NH systems.
Previous collision calculations on Rb+NH and Cs+NH
\cite{Tacconi:c:2007} did not include the effects of magnetic
fields and did not consider the spin-changing collisions that
are important for sympathetic cooling.

In this paper we assess the prospects for sympathetic cooling
of cold NH(X$^3\Sigma^-$) molecules with ultracold Li($^2S$)
atoms. We choose Li as the collision partner because its low
mass produces high centrifugal barriers that suppress
spin-changing collisions at low energies. $^7$Li has nuclear
spin $I_{\rm Li}=3/2$, and has a magnetically trappable
spin-stretched state with total angular momentum $F=2$ and
$M_F=2$. The $^{14}$NH molecule has complicated hyperfine
structure due to its electron spin $S=1$ and two nuclear spins
$I_{\rm N}=1$ and $I_{\rm H}=1/2$; its magnetically trappable
spin-stretched state has $F=5/2$ and $M_F=+5/2$. Collisions
between spin-stretched Li and NH have $M_S=m_{s_{\rm
Li}}+m_{s_{\rm NH}}=+3/2$ and thus occur entirely on the
quartet surface of Li-NH. We confine our attention to these
collisions. Collisions between states with $|M_S|\ne 3/2$
involve the doublet surface as well as the quartet surface and
are likely to lead to fast spin exchange.

The structure of this paper is as follows.
Section~\ref{sec:PES} describes electronic structure
calculations of the quartet potential energy surface for
Li($^2S$)+NH(X$^3\Sigma^-$). Section~\ref{sec:scat} describes
quantum scattering calculations and Section~\ref{sec:sym}
explores the prospects for sympathetic cooling.

\section{Potential Energy Surface}\label{sec:PES}

We have calculated the quartet interaction potential for
Li($^2S$)+NH(X$^3\Sigma^-$), using high-level \emph{ab initio}
electronic structure calculations. Calculations were performed
using an open-shell spin-restricted version \cite{Knowles:1993}
of the coupled cluster method \cite{Cizek:1966} with single,
double, and perturbative triple excitations [RCCSD(T)]. We used
the aug-cc-pVTZ basis set for N and H, and the uncontracted
cc-pVTZ basis for the Li atom \cite{Dunning:1989}, augmented by
Gaussian functions with exponents s:0.001, p:0.008, d:0.034,
and f:0.073. To improve the basis set for the Van der Waals
dispersion interaction we included spd bond functions (with
exponents 0.9, 0.3, 0.1 for s and p and 0.6 and 0.2 for d and
f) at the Li-NH center of mass. All \emph{ab initio}
calculations were performed using the MOLPRO package
\cite{molpro:2006} and corrected for basis-set superposition
error using the counterpoise method of Boys and Bernardi
\cite{Boys:1970}.

Calculations were performed on a grid in Jacobi coordinates
($R,\theta$), with a fixed NH bond length $r=1.0367$~\AA\
\cite{Brazier:1986}. $R$ is the separation between the NH
center of mass and the Li atom, and $\theta$ is the angle
between the NH bond vector and $R$, with $\theta=0$
corresponding to the linear NH-Li geometry. The angular grid is
a set of 11 angles corresponding to Gauss-Lobatto quadrature
points, and the radial grid extends from high up on the
repulsive potential wall to 15 \AA.

For each angle, radial interpolation  for $R\le15$~\AA\ and
extrapolation  for $R>15$~\AA\ is performed using the
reproducing kernel Hilbert space (RKHS) method
\cite{Ho:1996,Soldan:2000} to obtain the potential for any
given value of $R$. The potential is then expanded in Legendre
polynomials,
\begin{equation}
V(R,\theta)=\sum_\lambda V_\lambda(R)P_\lambda(\cos\theta),
\end{equation}
for $\lambda \le 8$. The radial strength coefficients $V_\lambda(R)$
are evaluated by integrating over the angular grid at each $R$ using
Gauss-Lobatto quadrature.

\begin{figure}
\begin{center}
\includegraphics[width=0.43\textwidth]{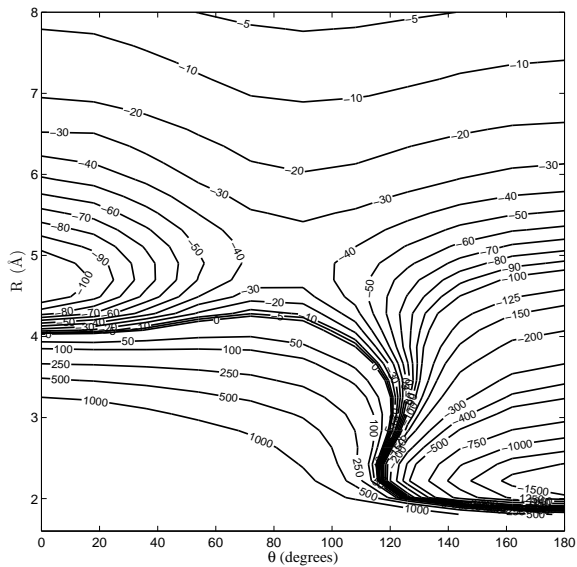}
\caption{Li($^2S$)+NH(X$^3\Sigma^-$) quartet potential
energy surface. Contours are labeled in cm$^{-1}$. $\theta=0^\circ$
corresponds to the linear N-H-Li geometry.}\label{fig:PES}
\end{center}
\end{figure}

Figure~\ref{fig:PES} shows the calculated quartet potential
energy surface for Li-NH. The global potential minimum is about
1800 cm$^{-1}$ deep and occurs at $\theta$=180$^\circ$ (Li-NH).
A secondary minimum about 113 cm$^{-1}$ deep occurs at
$\theta$=0$^\circ$. The large anisotropy is due to s-p mixing
of the Li orbitals, which is much stronger at the Li-NH
geometry than for NH-Li \cite{Soldan:MgNH:2009}.

Before examining the scattering dynamics, it is instructive to consider
some analytical properties of the potential. For a Van der Waals
potential that decays at long range as $-C_6/R^6$, a characteristic
potential length may be defined \cite{Jones:RMP:2006},
\begin{equation}
R_{\rm vdW} = \frac{1}{2}\left(\frac{2\mu C_6}{\hbar^2}\right)^{1/4}.
\end{equation}
Alternatively, the average scattering length $\bar{a}$, introduced by
Gribakin and Flambaum \cite{Gribakin:1993}, is
\begin{equation}
\bar{a}=\frac{4\pi}{\Gamma(1/4)^2} R_{\rm vdW} \approx 0.956R_{\rm vdW}.
\end{equation}
Gribakin and Flambaum showed that the scattering length follows
\begin{equation}
\frac{a}{\bar{a}}=1-\tan\left(\Phi-\frac{\pi}{8}\right),
\end{equation}
where
\begin{equation}
\Phi=\int_{R_0}^\infty \left[-2\mu V(R)/\hbar^2\right]^{1/2} dR
\end{equation}
is evaluated at threshold, with $R_0$ the inner classical turning
point. When $a$ is large and positive, the energy of the highest bound
state is just below threshold and is given by
\begin{equation}
E_{\rm bind} = -\frac{\hbar^2}{2\mu(a-\bar{a})^2}.
\end{equation}
If the potential is varied, by scaling or otherwise, the scattering
length passes through a pole every time an s-wave bound state occurs at
threshold. For Li($^2S$)+NH(X$^3\Sigma^-$), $C_6 = 164.9$ $E_{\rm
h}a_0^6$, and thus $R_{\rm vdW}=10.89$~\AA\ and $\bar{a}=$10.42~\AA,
corresponding to an average elastic cross section of $\bar{\sigma}_{\rm
elas}=4\pi \bar{a}^2=1365$~\AA$^2$.

\section{Scattering Calculations}\label{sec:scat}

\begin{figure}
\begin{center}
\includegraphics[width=0.45\textwidth]{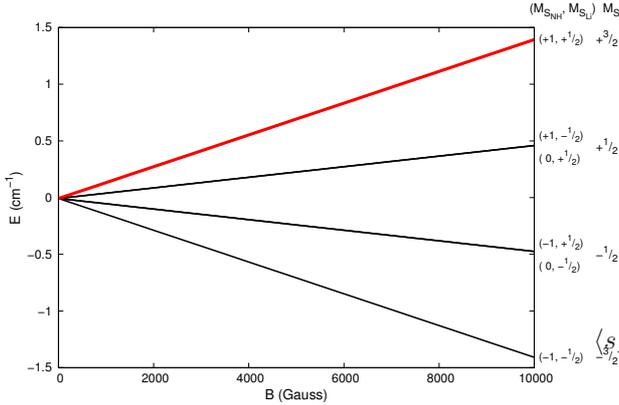}
\caption{(color online). Li($^2S$)+NH(X$^3\Sigma^-$) thresholds
as a function of magnetic field. The spin-stretched ($m_{s_{\rm
NH}}=+1, m_{s_{\rm Li}}=+\frac{1}{2}$) $M_{S}=+\frac{3}{2}$ state is
shown in red.}\label{fig:thres}
\end{center}
\end{figure}

The Hamiltonian for a $^2$S atom colliding with a rigid-rotor
$^3\Sigma$ diatomic molecule may be written
\begin{equation}\label{eqn:totham}
\hat{H} = -\frac{\hbar^2}{2\mu}\frac{d^2}{dR^2}+\frac{\hat{L}^2}{2\mu R^2}
+ \hat{H}_{\rm mon}
+ \hat{H}_{\rm Z} + V_{\rm ss}(R) + V(R,\theta),
\end{equation}
where $\hat{L}^2$ is the space-fixed end-over-end rotation
operator for the atom and the diatomic molecule about one
another, $\hat{H}_{\rm mon}$ is the field-free Hamiltonian for
the diatomic monomer, $\hat{H}_{\rm Z}$ represents the Zeeman
interaction, $V_{\rm ss}(R)$ is the anisotropic intermolecular
spin-spin interaction, $V(R,\theta)$ is the intermolecular
potential, and $\mu$ is the reduced mass of the colliding
system. The approximation of treating NH as a rigid rotor is
justified in more detail below.

The field-free monomer Hamiltonian for NH(X$^3\Sigma^-$) is
\begin{eqnarray}\label{eqn:monham}
\hat{H}_{\rm mon} &=& b_{\rm rot}\hat{n}^2
+ \gamma_{\rm ns}\hat{n}\cdot\hat{s_{\rm NH}}  \nonumber \\
&+& \lambda_{\rm ss}\left[\frac{4\pi}{5}\right]^{\frac{1}{2}}
\sqrt{6}\sum_q(-1)^qY_{2-q}(\mathbi{r})
\left[s_{\rm NH}\otimes s_{\rm NH}\right]^{(2)}_q,
\end{eqnarray}
where $\hat{n}$ and $\hat{s}_{\rm NH}$ are the NH rotation and
electron spin angular momentum operators, $b_{\rm
rot}=\hbar^2/2\mu r_{\rm mon}^2 =$ 16.343 cm$^{-1}$ is the NH
rotational constant, $\gamma_{\rm ns}= -0.055$ cm$^{-1}$ is the
spin-rotation constant and $\lambda_{\rm ss} = 0.92$ cm$^{-1}$
is the spin-spin interaction constant. The values of the
spectroscopic constants for the vibrational ground state are
from ref.\ \cite{Mizushima}. The Zeeman Hamiltonian is
\begin{equation}
\hat{H}_{\rm Z} = g_e\mu_{\rm B} \hat{B}\cdot(\hat{s}_{\rm NH}+\hat{s}_{\rm Li}),
\end{equation}
where $\hat{B}$ is the magnetic field vector, $\mu_{\rm B}$ is
the Bohr magneton and $g_e$ is the electron $g$-factor. The
anisotropic intermolecular spin-spin interaction can be
expressed  as
\begin{equation}
V_{\rm ss}(R) = \lambda(R)\left[\hat{s}_{\rm Li}\cdot\hat{s}_{\rm NH}
-3(\hat{s}_{\rm Li}\cdot\hat{e}_{\rm R})(\hat{s}_{\rm NH}\cdot\hat{e}_{\rm R})\right],
\end{equation}
where $\hat{s}_{\rm Li}$ and $\hat{s}_{\rm NH}$ are the spin
angular momentum operators for the Li atom and the NH molecule
respectively, $\hat{e}_{\rm R}$ is a unit vector along $R$, and
$\lambda(R)=-E_{\rm h}\alpha^2/(R/a_0)^3$, where
$\alpha\approx1/137$ is the fine-structure constant.

We construct the collision Hamiltonian in the fully uncoupled
basis set $|n m_n\rangle |s_{\rm NH} m_{s_{\rm
NH}}\rangle|s_{\rm Li} m_{s_{\rm Li}}\rangle|L M_L\rangle$,
where $m_i$ is the projection of the angular momentum term
$s_i$ onto the space-fixed magnetic field axis. The matrix
elements for NH monomer operators are the same as for
scattering of NH from a closed-shell atom
\cite{Gonzalez-Martinez:2007}, with the addition of factors
$\delta_{m_{s{\rm A}}{m_{s{\rm A}}'}}$. The intermolecular
spin-spin interaction has matrix elements
%\begin{widetext}
\begin{eqnarray}\hspace{-1.1cm}
\langle s_{\rm A} m_{s{\rm A}} s_{\rm B} m_{s{\rm B}} n_{\rm B}  m_{n{\rm B}} L M_L |V_{\rm ss}|
 s_{\rm A} m'_{s{\rm A}} s_{\rm B} m'_{s{\rm B}} n'_{\rm B}  m'_{n{\rm B}} L' M'_L \rangle \nonumber\\
= \sqrt{30} \lambda(R) \delta_{n_{\rm B} n'_{\rm B}}
\delta_{m_{n {\rm B}} m'_{n {\rm B}}}
(-1)^{ s_{\rm A} + s_{\rm B} - m_{s{\rm A}} -  m_{s{\rm B}} -M_L } \nonumber\\
\left[s_{\rm A}(s_{\rm A}+1)(2s_{\rm A} +1) s_{\rm B} (s_{\rm B} +1) (2s_{\rm B} +1)(2L+1)(2L'+1)\right]^\frac{1}{2}
\nonumber\\
\left( \begin{array}{ccc}  L  & 2  & L' \\  0 & 0 & 0  \end{array} \right)
\sum_{q_1 q_2}
\left( \begin{array}{ccc}  L  & 2  & L' \\  -M_L & -q_1-q_2 &  M'_L  \end{array} \right) \nonumber\\
\left( \begin{array}{ccc}  1  & 1  & 2  \\  q_1 & q_2 & -q_1-q_2  \end{array} \right)
\left( \begin{array}{ccc}   s_{\rm A}   &  1   &   s_{\rm A} \\  -m_{s{\rm A}}  &  q_1   & m'_{s{\rm A}}   \end{array} \right)
\left( \begin{array}{ccc}   s_{\rm B}   &  1   &   s_{\rm B} \\  -m_{s{\rm B}}  &  q_2   & m'_{s{\rm B}}   \end{array} \right).
\end{eqnarray}

The total spin $S$ of a system made up of an open-shell atom
and an open-shell molecule can take values between $|s_{\rm A}
- s_{\rm B}|$ and $s_{\rm A}+s_{\rm B}$. For Li--NH the allowed
values are $S=\frac{1}{2}$ (doublet) and $\frac{3}{2}$
(quartet). The interaction potential $ V_{\rm int} (R,\theta)$
may be written in terms of projection operators,
\begin{equation}
  V_{\rm int} (R,\theta) = \sum_{S=-| s_{\rm A} + s_{\rm B} |  }^{s_{\rm A} + s_{\rm B} }
 |S \rangle V_S(R,\theta) \langle S |
\end{equation}
and the matrix element of  $ V_{\rm int} (R,\theta)$ is
\begin{eqnarray}\hspace{-1.1cm}
\langle s_{\rm A} m_{s{\rm A}} s_{\rm B} m_{s{\rm B}} n_{\rm B}  m_{n {\rm B}} L M_L |V_{\rm int} (R,\theta)|
 s_{\rm A} m'_{s_{\rm A}} s_{\rm B} m'_{s{\rm B}} n'_{\rm B}  m'_{n {\rm B}} L' M'_L \rangle
\nonumber \\
= \sum_S
(-1)^{2s_{\rm A} + 2 s_{\rm B} -  m_{s{\rm A}} -  m_{s{\rm B}} - M_L  }  (2S+1) \nonumber\\
 \langle n_{\rm B}  m_{n{\rm B}} L M_L  |  V_S (R,\theta) | n'_{\rm B}  m'_{n{\rm B}} L' M'_L  \rangle \nonumber \\
\left( \begin{array}{ccc}  s_{\rm A}& s_{\rm B}& S \\ m_{s{\rm A}} & m_{s{\rm B}} & -m_{s{\rm A}}-m_{s{\rm B}} \end{array}
   \right)
\left( \begin{array}{ccc}  s_{\rm A}& s_{\rm B}& S \\ m'_{s{\rm A}}&  m'_{s{\rm B}}&  -m'_{s{\rm A}}-m'_{s{\rm B}} \end{array}
   \right).
\label{intpot}
\end{eqnarray}
%\end{widetext}
The doublet and quartet interaction potentials $V_S(R,\theta)$
have the same long-range coefficients, so become degenerate
when the N atom and NH molecule are far enough apart that their
valence shells do not overlap. In the present work we
approximate the operator $V_{\rm int} (R,\theta)$ by taking
$V_S=V_{3/2}$ for both spin states. This approximation is
reasonable because we wish to focus on collisions between atoms
and molecules in spin-stretched states, and $V_{1/2}$ has no
matrix elements (diagonal or off-diagonal) involving such
states. When this approximation is made, orthogonality
relations for the $3j$ symbols reduce Eq.\ \ref{intpot} to a
form diagonal both in $m_{s{\rm A}}$ and $m_{s{\rm B}}$. The
explicit expression for $\langle n_{\rm B} m_{n{\rm B}} L M_L |
V_S (R,\theta) | n_{\rm B}  m'_{n{\rm B}} L' M'_L  \rangle $ is
the same as for scattering of NH from a closed-shell atom
\cite{Gonzalez-Martinez:2007}, with the addition of factors
$\delta_{m_{s{\rm A}}{m_{s{\rm A}}'}}$.

In a magnetic field, the lowest Li-NH threshold ($N=0,s_{\rm
NH}=1,s_{\rm Li}=\frac{1}{2}$) splits into six Zeeman
sub-levels, as shown in Figure~\ref{fig:thres}. There is one
$M_S=-\frac{3}{2}$ state, two degenerate $M_S=-\frac{1}{2}$
states, two degenerate $M_S=\frac{1}{2}$ states, and a single
spin-stretched $M_S=\frac{3}{2}$ state.

In a magnetic field the projection of the total angular
momentum $M_{\rm tot}=m_n+M_S+M_L$ is conserved. Thus in the
rotational ground state $N=0$, any change in $M_S$ must be
accompanied by a change in $M_L$. For s-wave scattering ($L=0$)
from the spin-stretched $M_S=+\frac{3}{2}$ state, this requires
that $L=0\rightarrow L\ge2$. Thus the dominant outgoing
channels are $|M_S=\pm\frac{1}{2},L=2\rangle$ and
$|M_S=-\frac{3}{2},L=4\rangle$, which have centrifugal barriers
that suppress inelastic scattering at low energies and magnetic
fields. The heights of the centrifugal barriers are
approximately $E_{\rm cf}^L = (\hbar^2
L(L+1)/\mu)^{\frac{3}{2}}(54C_6)^{-\frac{1}{2}}$, which for
LiNH are $E_{\rm cf}^2 =60$ mK and $E_{\rm cf}^4=368$ mK.

%and thus for s-wave scattering ($L=0$) from the spin-stretched $M_S=+\frac{3}{2}$ state there are no outgoing channels with $L=0$ and thus the dominant outgoing channels are the $|M_S=\pm\frac{1}{2},L=2\rangle$ and the $|M_S=-\frac{3}{2},L=4\rangle$ channels, which are suppressed at low energies by centrifugal barriers. The heights of the centrifugal barriers are approximately $E_{\rm cf}^L = (\hbar L(L+1/\mu)^{\frac{3}{2}}(54C_6)^{-\frac{1}{2}}$, which for LiNH are $E_{\rm cf}^2 =60$ mK and $E_{\rm cf}^4=368$ mK.

For magnetically trapped spin-stretched states, collisional spin
relaxation to other Zeeman states will result in heating and/or loss of
molecules from the trap. Spin relaxation occurs via two mechanisms. The
first is via the direct coupling of the incident state to outgoing
states through the anisotropic spin-spin interaction, by a similar
mechanism to spin relaxation in collisions between alkali-metal atoms.
The second occurs via the interplay of the potential anisotropy and the
intramolecular NH spin-spin interaction \cite{Krems:mfield:2004}, which
results in an inelastic cross section proportional to $1/b_{\rm rot}^2$
\cite{Cybulski:2005,Campbell:2009}.

We have carried out low-energy quantum scattering calculations
using the MOLSCAT package \cite{molscat:v14}, as modified to
handle collisions in magnetic fields
\cite{Gonzalez-Martinez:2007}. Sets of coupled-channel
equations are constructed for each $M_{\rm tot}$ in the
uncoupled basis set described above. Solutions to the
coupled-channel equations are then propagated over a grid in
$R$, using the hybrid log-derivative method of Alexander and
Manolopoulos \cite{Alexander:1987}, which in the short-range
region ($R_{\rm min}\le R < R_{\rm mid}$) uses a
fixed-step-size log-derivative propagator and in the long-range
region ($R_{\rm mid}\le R \le R_{\rm max}$) a variable
step-size Airy propagator. At $R_{\rm max}$, the log-derivative
solutions are transformed into the basis set of asymptotic
eigenfunctions and matched to scattering boundary conditions to
obtain the $S$-matrix $S^{\rm M_{\rm tot}}$. The scattering
cross sections between levels $i$ and $f$ are then obtained
from
\begin{equation}
\sigma_{i\rightarrow f} = \frac{\pi}{k_{i}^2}\sum_{M_{\rm tot}L_iL_f}
\left|\delta_{if}\delta_{L_iL_f}-S^{M_{\rm tot}}_{iL_ifL_f}\right|^2,
\end{equation}
where $k_{i}$ is the wave vector in the incident channel,
$k_i^2=2\mu(E-E_i)/\hbar^2$.

\begin{figure}
\begin{center}
\includegraphics[width=0.49\textwidth]{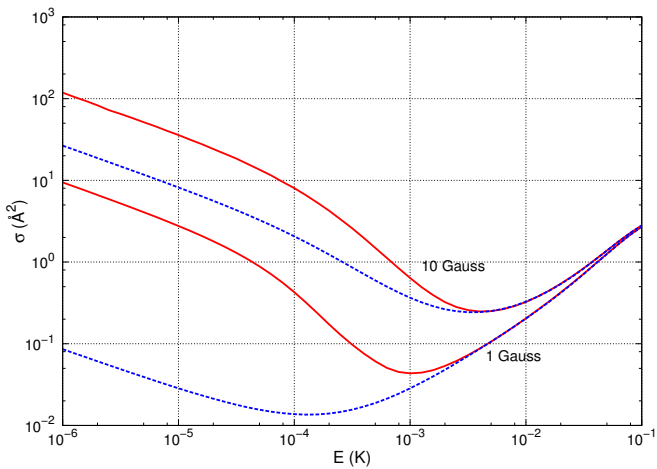}
\caption{(color online). S-wave total inelastic cross sections calculated
as a function of collision energy for $R_{\rm AX} > R_{\rm max}=50$ \AA
~(blue, dashed) and $R_{\rm AX} < R_{\rm max}=600$ \AA ~(red, solid)
for magnetic fields of 1 and 10 Gauss. $R_{\rm AX}$(1 G) = 476 \AA~ and
$R_{\rm AX}$(10 G) = 150 \AA. }\label{fig:shortrange}
\end{center}
\end{figure}

The anisotropic intermolecular spin-spin interaction directly
couples the incident s-wave channel $M_S=+\frac{3}{2},L=0$ to
the two $M_S=+\frac{1}{2},L=2$ channels and to the
$(-1,+\frac{1}{2}), M_S=-\frac{1}{2},L=2$ channel. Since the
$L=2$ channels have barriers at long range, there are narrowly
avoided crossings between the channel adiabats
\cite{Janssen:NHNH:field:2011} at very long range when the
kinetic energy release is small (at low magnetic field). If the
long-range behaviour in both the $L=0$ and $L=2$ channels is of
the form
\begin{equation}
V(R) \xrightarrow[R\rightarrow\infty]{} \frac{\hbar^2 L(L+1)}{2\mu R^2}
- \frac{C_6}{R^6} +\mathcal{O}(R^{-7}),
\end{equation}
these avoided crossings are at a distance $R_{\rm AX}$ such
that
\begin{equation}
g_e\mu_{\rm B}B\Delta M_S = \frac{\hbar^2 \left[L_f(L_f+1)-L_i(L_i+1)\right]}{2\mu R_{\rm AX}^2},
\end{equation}
where $g_e\mu_{\rm B}B\Delta M_S$ is the energy separation of
the two adiabats due to the magnetic field $B$, $L_i$ and $L_f$
are the values of $L$ in the incoming and outgoing channels,
and $\Delta M_S=M_{S_f}-M_{S_i}$. To account correctly for the
intermolecular spin-spin coupling, the scattering solutions
need to be propagated out beyond the largest value of $R_{\rm
AX}$. Figure~\ref{fig:shortrange} shows the effect of placing
$R_{\rm max}<R_{\rm AX}$ and $R_{\rm AX}<R_{\rm max}$ on the
total inelastic s-wave cross sections as a function of
collision energy. We consider only magnetic fields relevant to
sympathetic cooling down to 1~G, at which point the largest
$R_{\rm AX}=476$~\AA, and thus the propagation parameters
$R_{\rm min}=$ 1.8 \AA, $R_{\rm mid}=12.5$~\AA\ and $R_{\rm
max}=600$~\AA\ were used throughout.

\begin{figure}
\begin{center}
\includegraphics[width=0.49\textwidth]{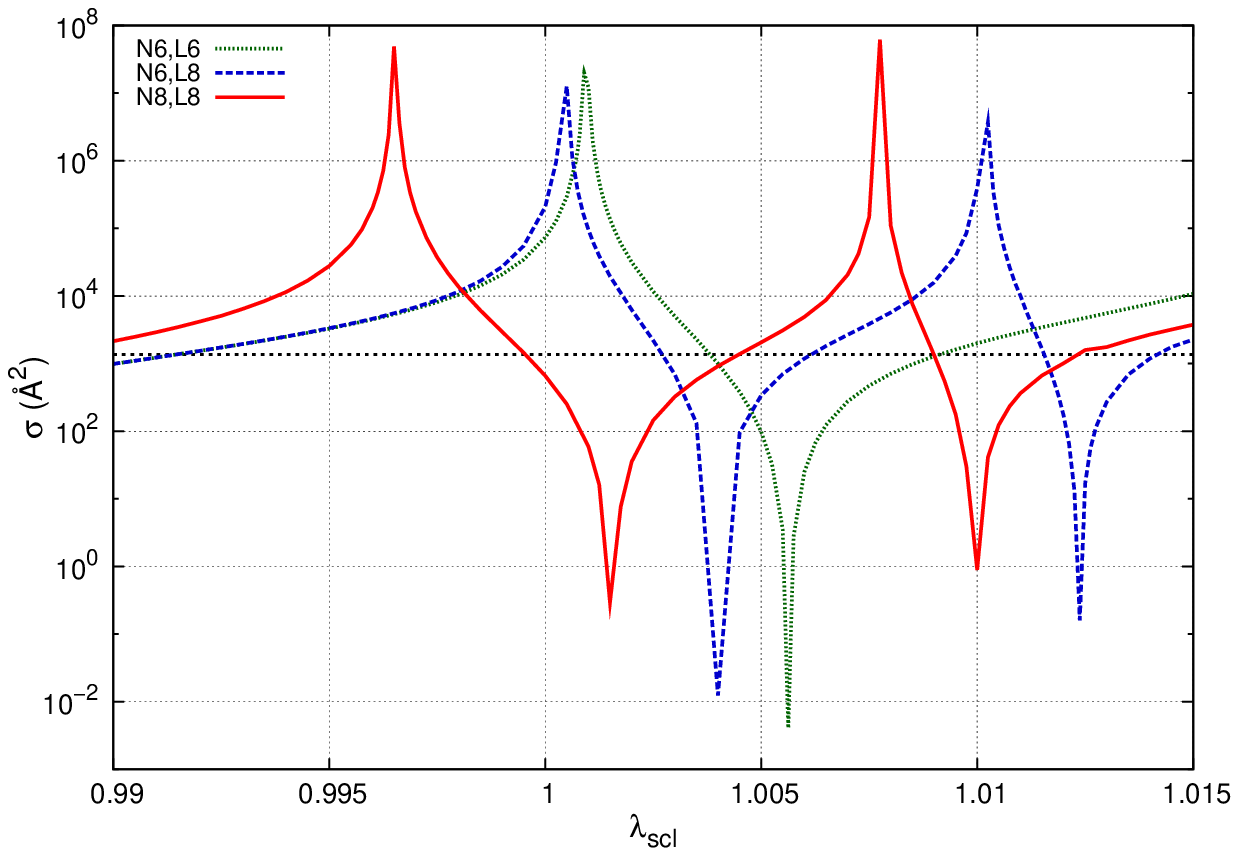}
\includegraphics[width=0.49\textwidth]{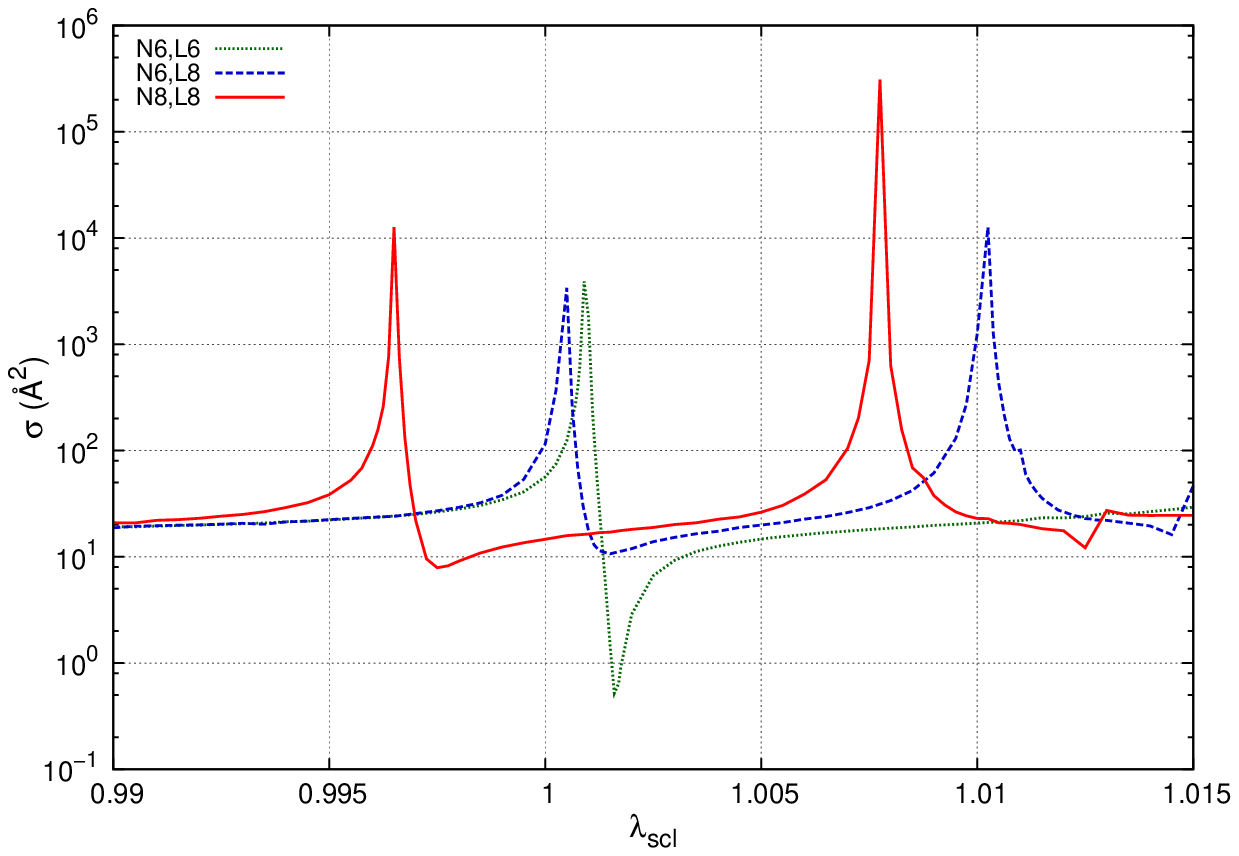}

\caption{(color online). Spin-stretched s-wave elastic (upper
panel) and inelastic (lower panel) cross sections calculated
for basis sets with $n_{\rm max}=6,L_{\rm max}=6$ (green,
dotted), $n_{\rm max}=6,L_{\rm max}=8$ (blue, dashed), and
$n_{\rm max}=8,L_{\rm max}=8$ (red, solid), as a function of
the potential scaling factor $\lambda_{\rm scl}$ at a collision
energy of $10^{-6}$~K and a magnetic field strength of 10
Gauss. The average cross section $\bar{\sigma}_{\rm
elas}=4\pi\bar{a}^2=1365$~\AA$^2$ is also shown (black,
heavy-dots).}\label{fig:basisconv}
\end{center}
\end{figure}

Low-energy scattering depends strongly on the details of the
potential energy surface. Thus when converging the basis set
for the scattering calculations we need consider the
uncertainty in the interaction potential as well as convergence
with respect to the variables $n_{\rm max}$ and $L_{\rm max}$
that govern the basis set size
\cite{Janssen:NHNH:field-free:2011}. To explore the convergence
with respect to the potential we introduce a scaling factor
$\lambda_{\rm scl}$ that allows the interaction potential to be
varied within its error bounds,
\begin{equation}
V^{\rm scaled}(R,\theta)=\lambda_{\rm scl}V(R,\theta).
\end{equation}
Since the main features of the cross sections are determined by
the scattering length and the positions of bound states that
cause resonances, both of which are tuned by $\lambda_{\rm
scl}$, scaling the potential in this way is sufficient to
explore the range of possible collision behaviour. This also
justifies fixing the NH bond length $r$ in our calculations:
the collisions are approximately vibrationally adiabatic, and
allowing $r$ to relax would simply deepen the potential
slightly and correspond approximately to a value of
$\lambda_{\rm scl}>1$.

Figure~\ref{fig:basisconv} shows the potential-dependence of
the s-wave spin-stretched elastic scattering cross section for
the three basis sets $n_{\rm max}=6,L_{\rm max}=6$ (645
channels), $n_{\rm max}=6,L_{\rm max}=8$ (937 channels), and
$n_{\rm max}=8,L_{\rm max}=8$ (1403 channels) at a collision
energy of $10^{-6}$ K and a magnetic field of 10 G.
Calculations with $n_{\rm max}=8,L_{\rm max}=10$ (1887
channels) gave results essentially identical to those with
$n_{\rm max}=8,L_{\rm max}=8$. The computer time scales as the
number of channels cubed. It may be seen that, for all three
basis sets, the elastic cross sections fluctuate about the
average cross section $\bar{\sigma}_{\rm elas}$ and the total
inelastic cross sections have peaks at corresponding positions.
The fluctuations (resonances) have quite similar form for
different basis sets, but the positions of the resonances are
shifted. This is similar to the result obtained by Janssen {\em
et al.} \cite{Janssen:NHNH:field-free:2011} for NH-NH.

\begin{figure}
\begin{center}
\includegraphics[width=0.49\textwidth]{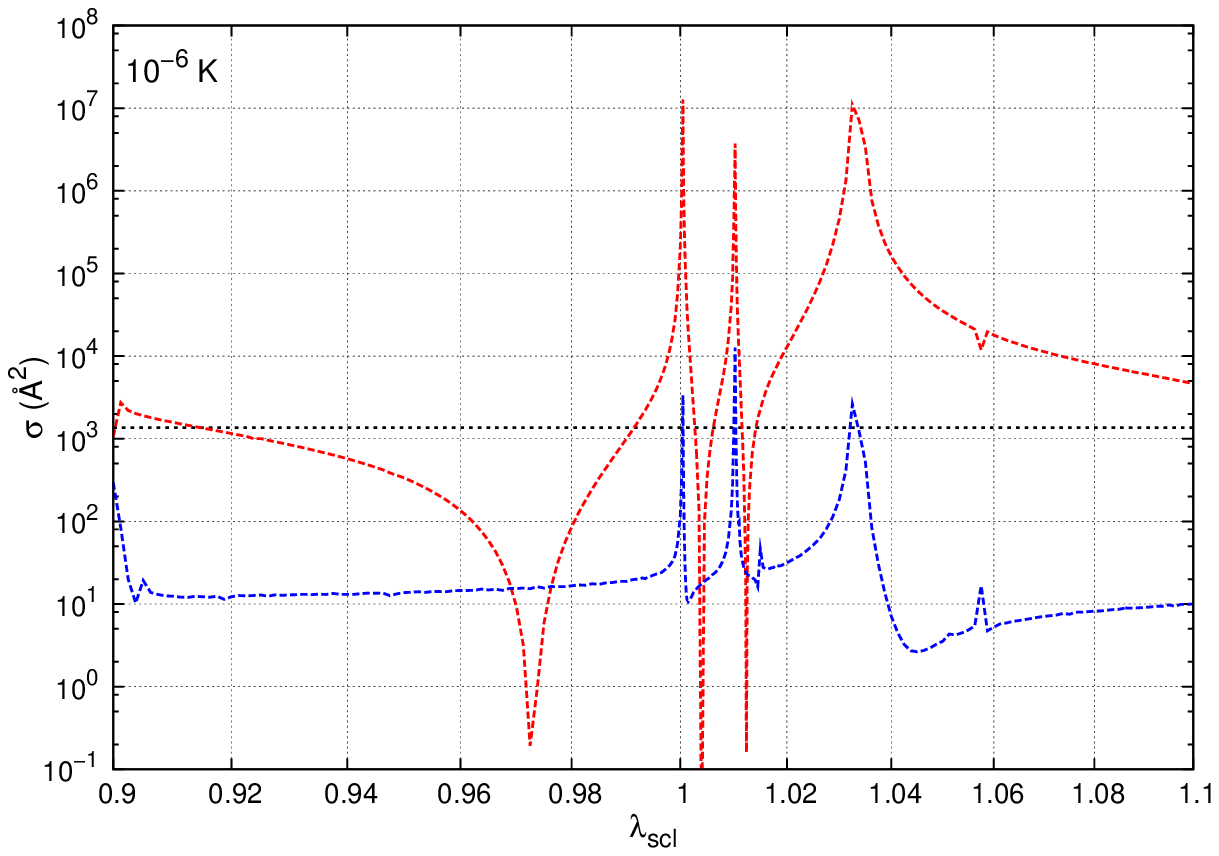}
\includegraphics[width=0.49\textwidth]{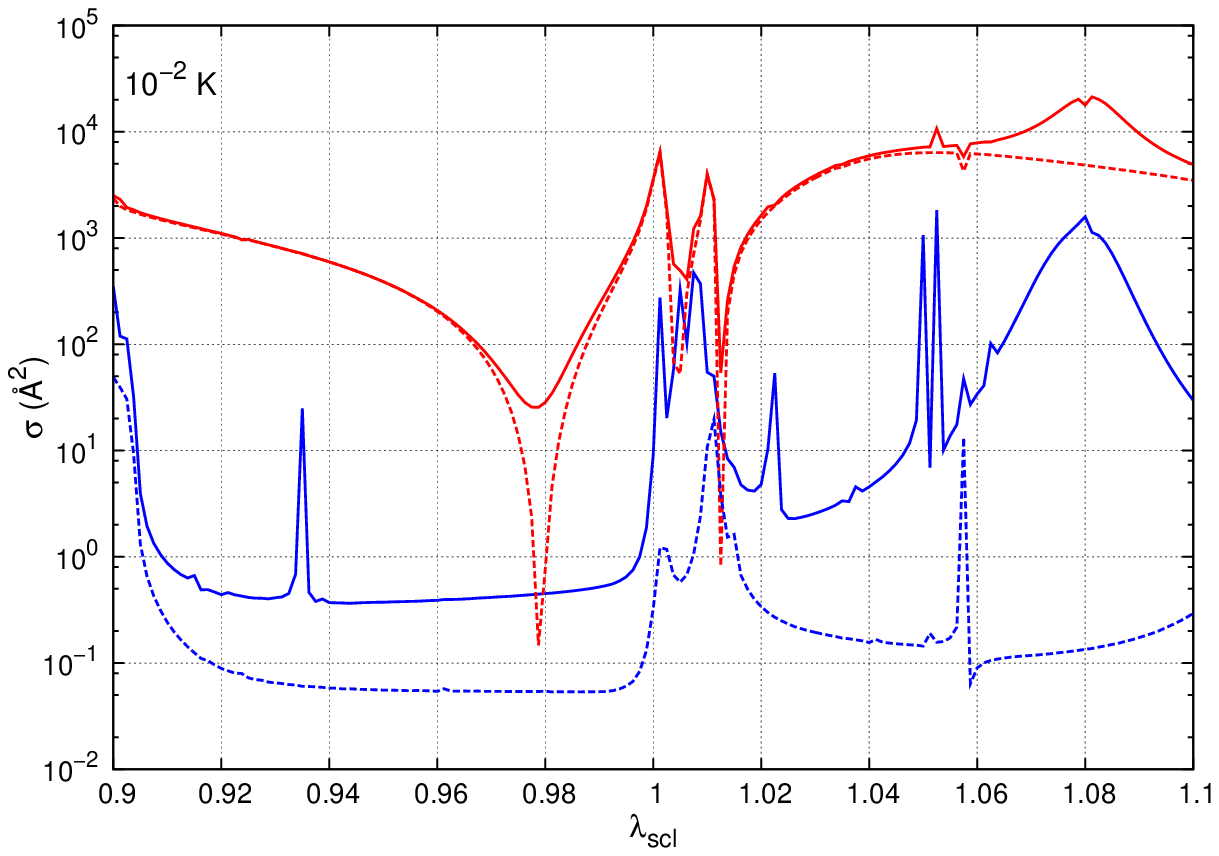}
\includegraphics[width=0.49\textwidth]{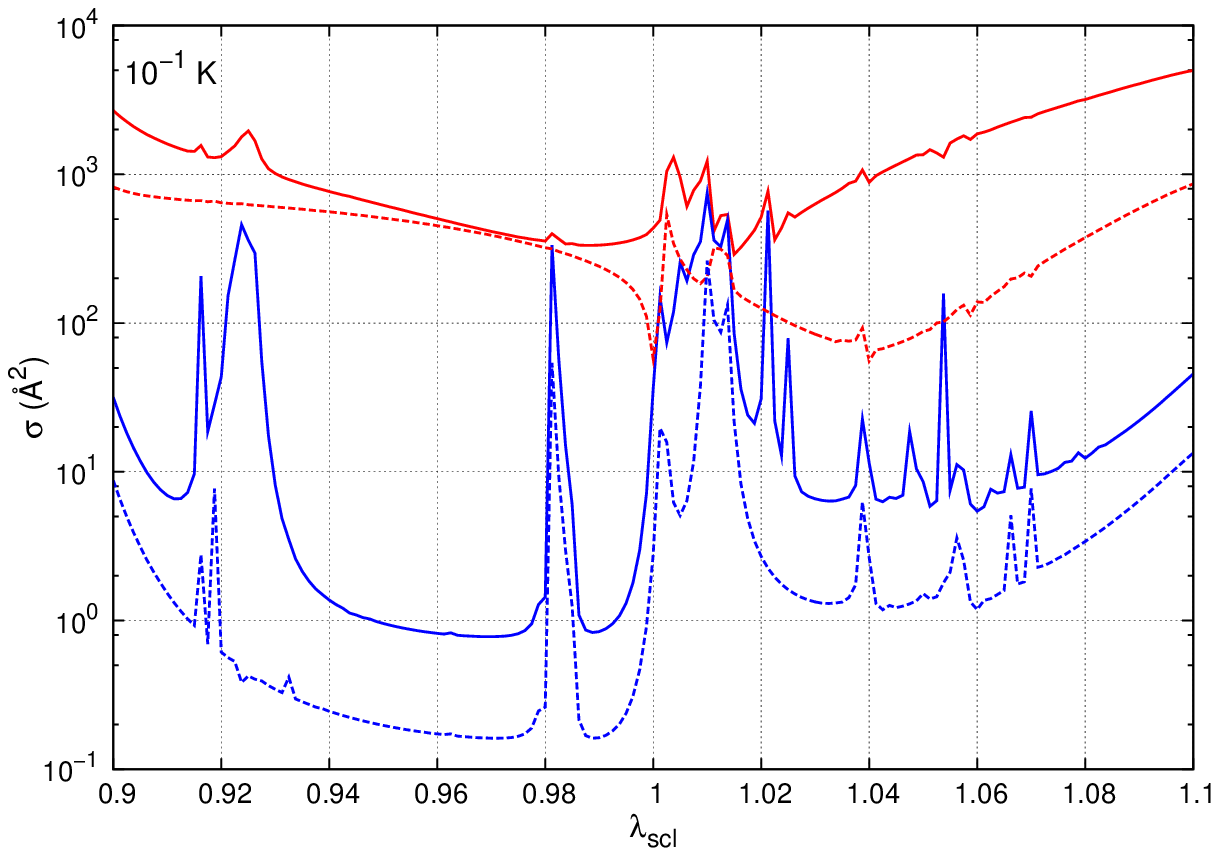}
\caption{(color online).
Dependence of the elastic (red) and total inelastic (blue)
cross sections on the potential scaling factor $\lambda_{\rm
scl}$ at collision energies of 1~$\mu$K (top), 10~mK (center)
and 100~mK (bottom) at a magnetic field strength of 10~G. The
s-wave contribution to each cross section is shown as a dashed
line (and is the only significant contribution at $10^{-6}$~K).
The average s-wave cross section $\bar{\sigma}_{\rm
elas}=4\pi\bar{a}^2=$ 1365 \AA$^2$ is shown as a dotted black
line. The calculations use the basis set with $n_{\rm
max}=6,L_{\rm max}=8$.}\label{fig:potdep}
\end{center}
\end{figure}

It is clear that basis sets smaller than $n_{\rm max}=8,L_{\rm
max}=8$ do not give ``converged" results for a specific
interaction potential. However, the calculated interaction
potential is itself probably accurate only to about $\pm5$\%.
Even a fully converged basis set would therefore not give
reliable predictions of numerical values for elastic (or
inelastic) cross sections in the ultracold regime.
Nevertheless, {\em even an unconverged basis set} reliably
predicts the {\em probability} that the elastic and inelastic
cross sections will lie in a particular range of values, and
thus the likelihood that sympathetic cooling can succeed. Until
the actual cross sections are measured, we have to accept this
uncertainty in the results, and until then there is no point in
using very expensive basis sets that aim for complete
convergence.

The converged $n_{\rm max}=8,L_{\rm max}=8$ basis is
computationally too expensive for calculations involving p-wave
and higher partial-wave contributions, which are required at
temperatures above 10 $\mu$K. The smaller and more
computationally manageable $n_{\rm max}=6,L_{\rm max}=8$ has
therefore been used in the remainder of this work.
Figure~\ref{fig:potdep} shows the dependence of the elastic and
total inelastic cross sections with this basis set for a wider
range of $\lambda_{\rm scl}$ for collision energies of
1~$\mu$K, 10~mK and 100~mK: it may be seen that the higher
partial waves that contribute at higher energies ``fill in" the
deep dips that occur in the s-wave cross sections, but that
additional structure appears at higher energy because of
resonances in higher partial waves.

\section{Prospects for Sympathetic Cooling}\label{sec:sym}

\begin{figure}
\begin{center}
\includegraphics[width=0.49\textwidth]{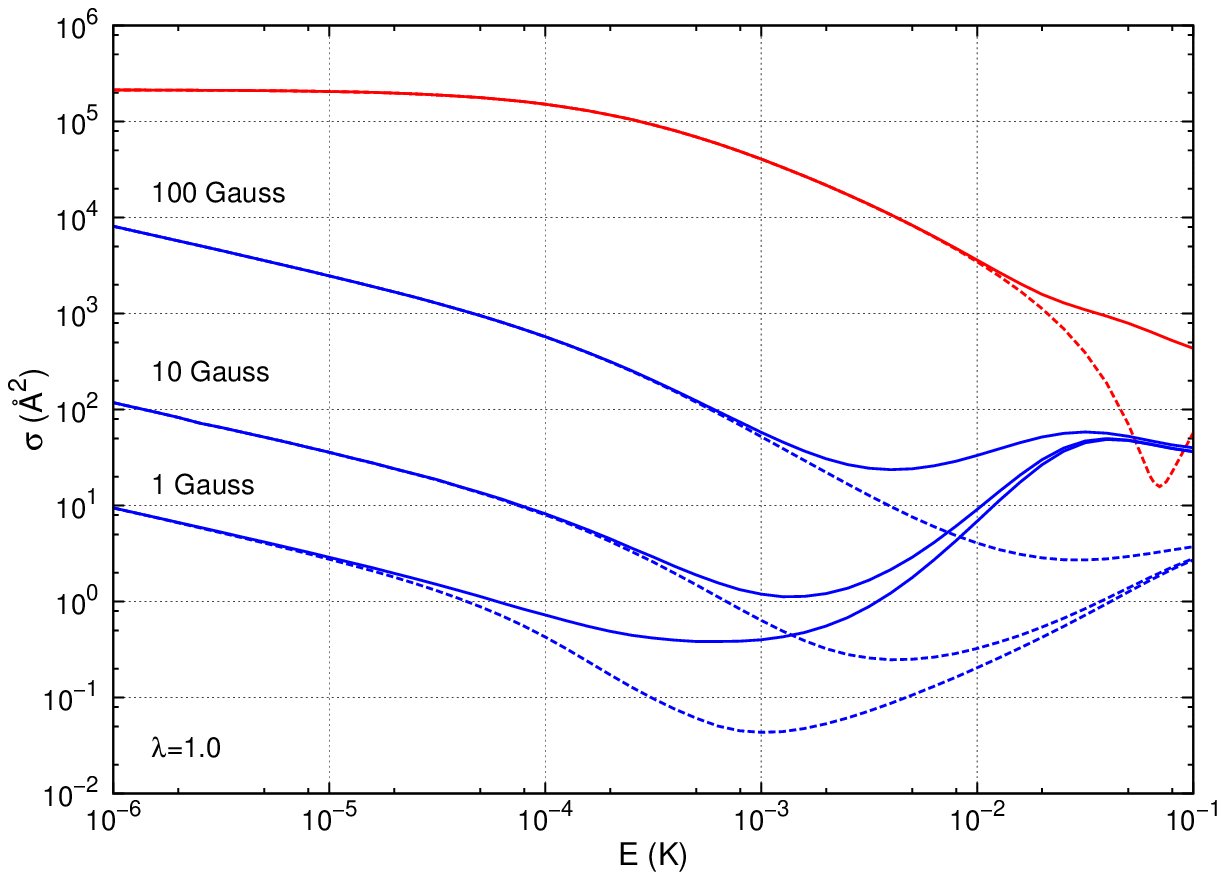}
\includegraphics[width=0.49\textwidth]{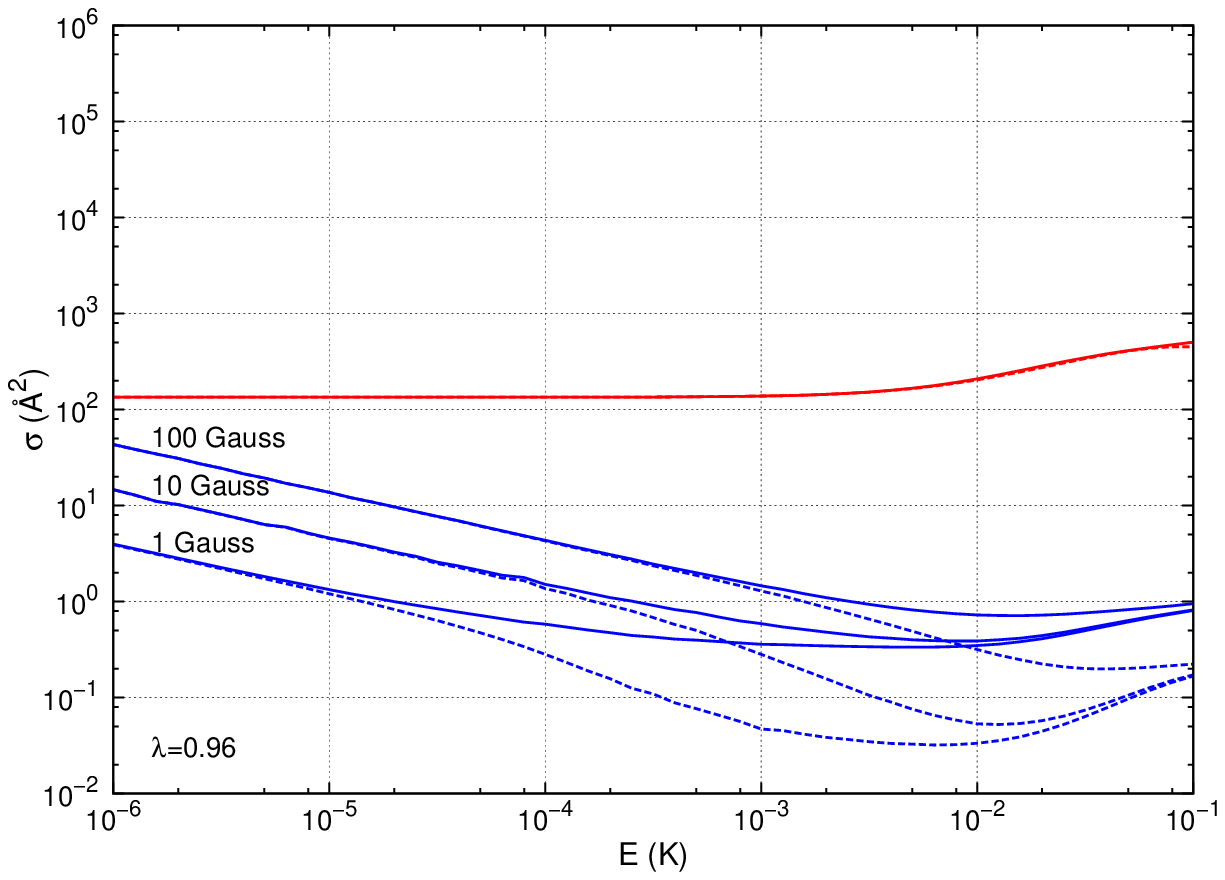}
\includegraphics[width=0.49\textwidth]{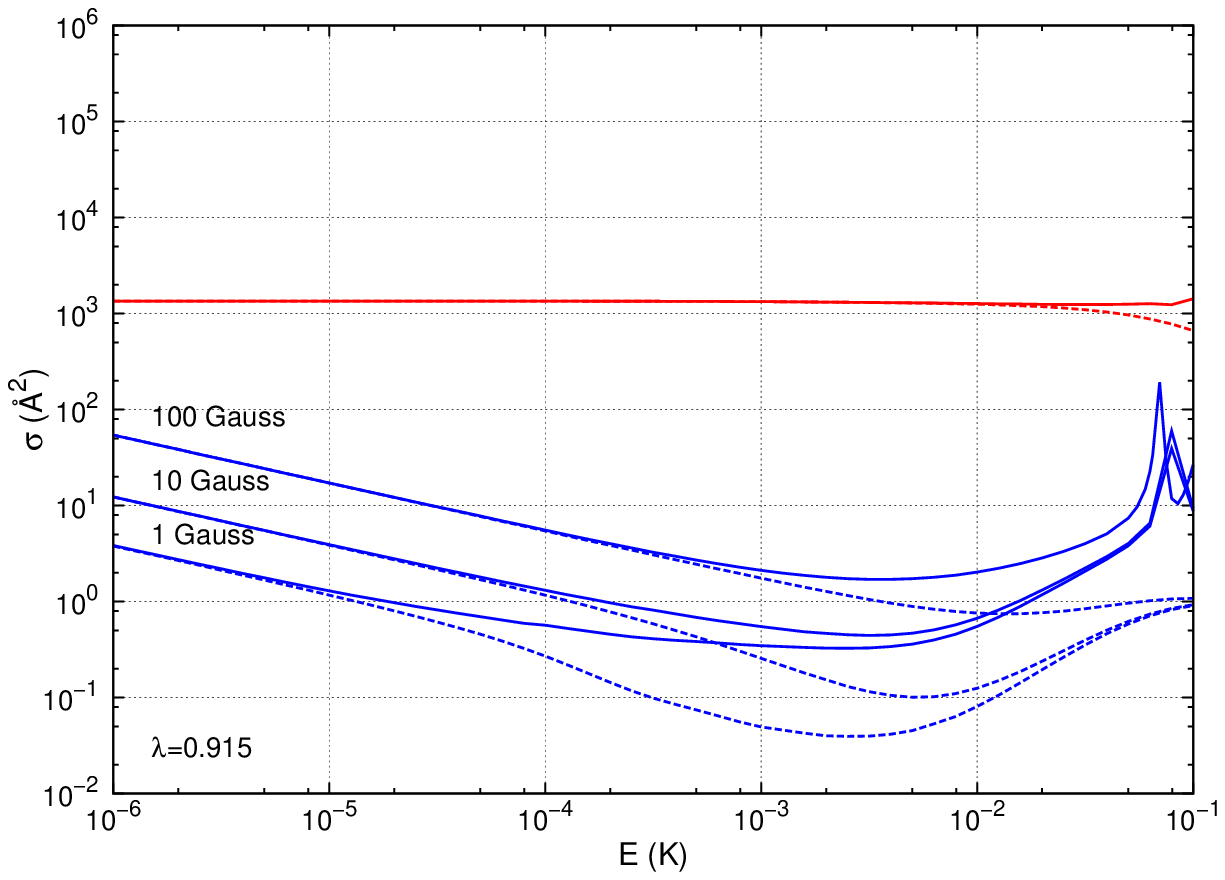}
\caption{(color online). Li+NH spin-stretched elastic (solid,
red) and total inelastic (solid, blue) cross sections as a
function of collision energy for various magnetic fields, for
the potentials with $\lambda_{\rm scl}=1$, 0.96 and 0.915,
corresponding to $a=-12.5\overline{a}$, $0.3\overline{a}$,
$\overline{a}$, respectively. The cross sections include all s,
p, and d-wave ($L=0,1,2$) contributions. The s-wave elastic
(dashed,red) and total inelastic (dashed, blue) cross sections
are also shown.}\label{fig:L1Edep}
\end{center}
\end{figure}

\begin{figure*}[t]
\begin{center}
\subfigure{\includegraphics[width=0.3\textwidth]{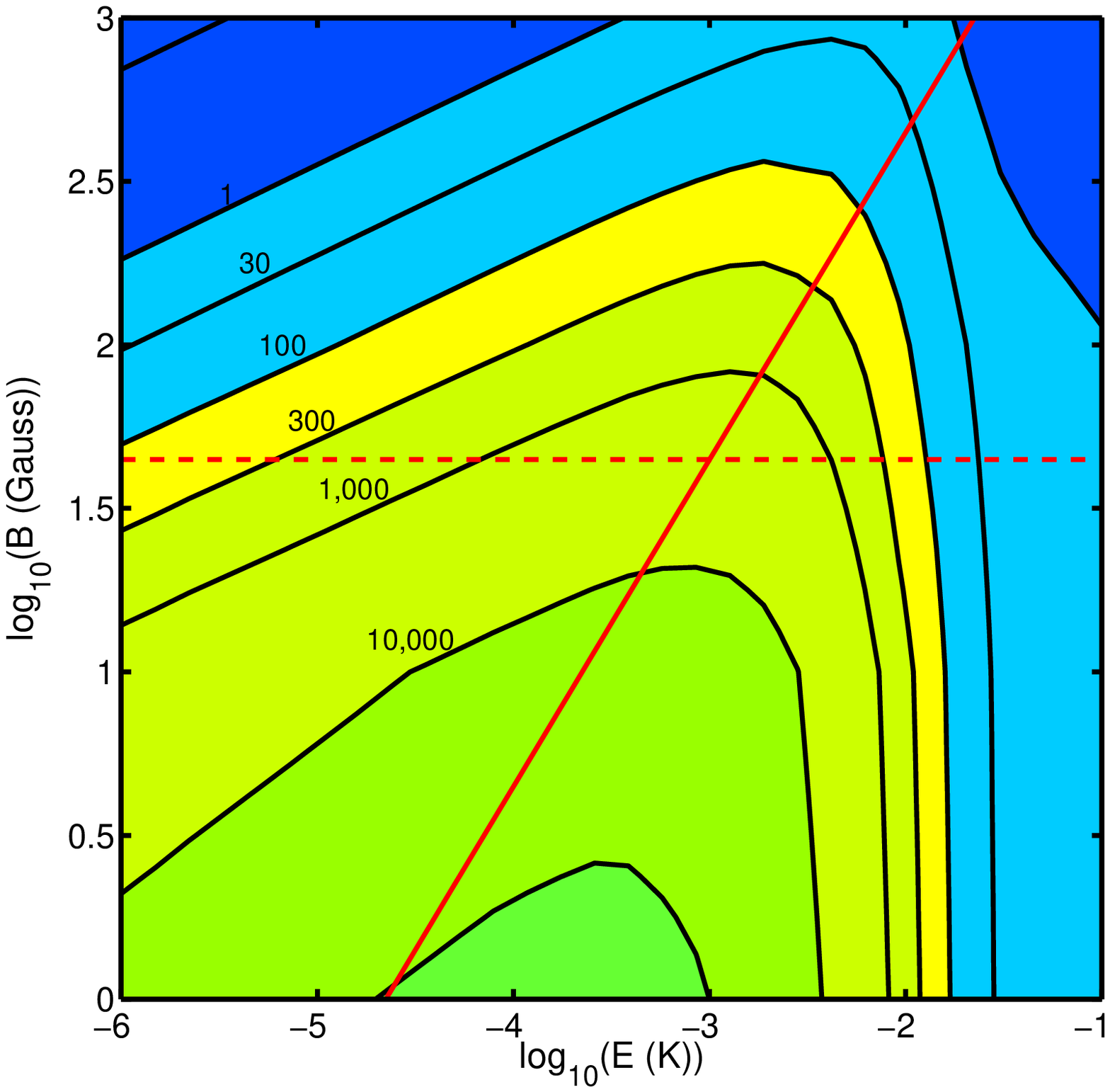}}
\qquad
\subfigure{\includegraphics[width=0.3\textwidth]{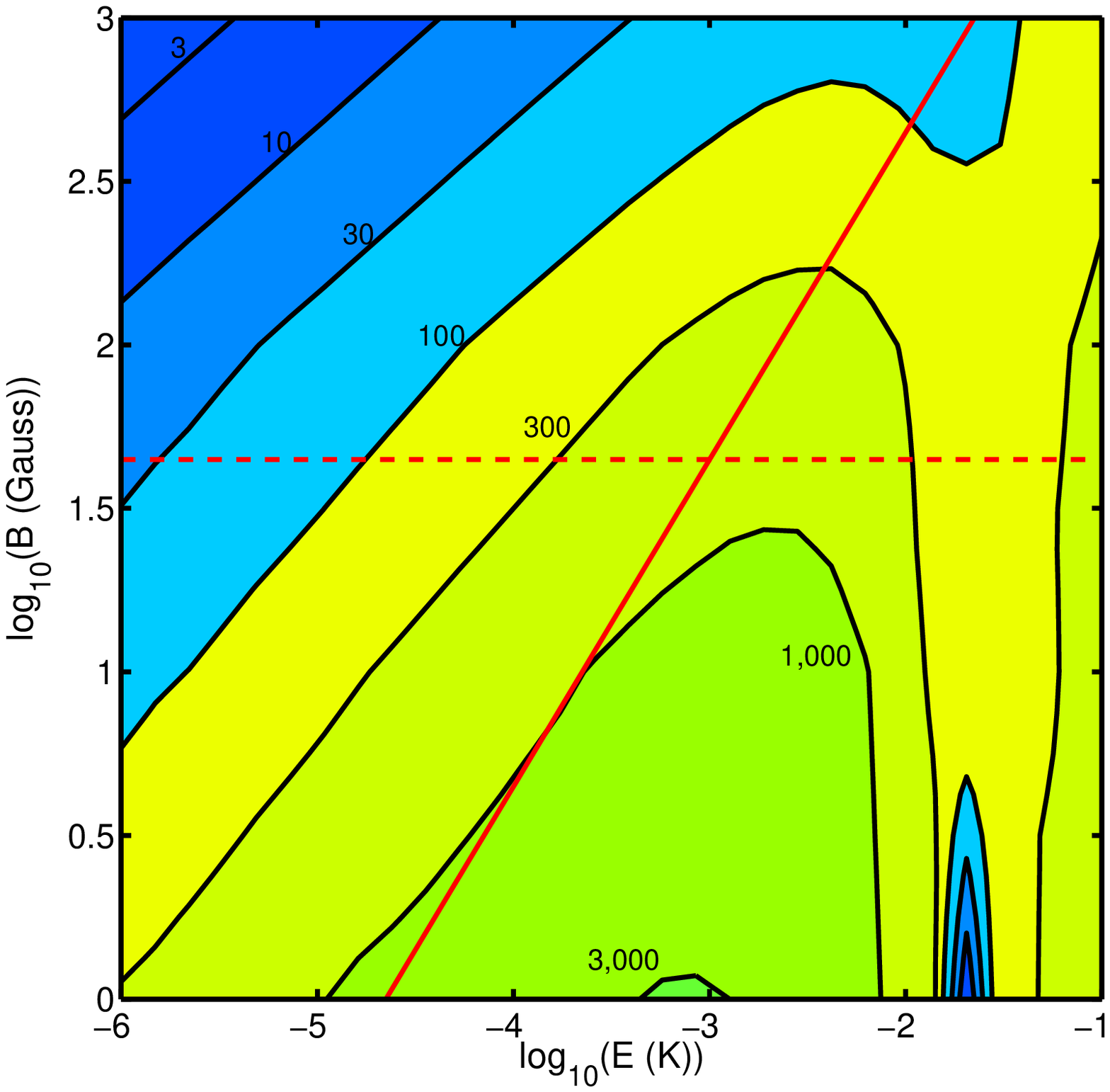}}
\qquad
\subfigure{\includegraphics[width=0.3\textwidth]{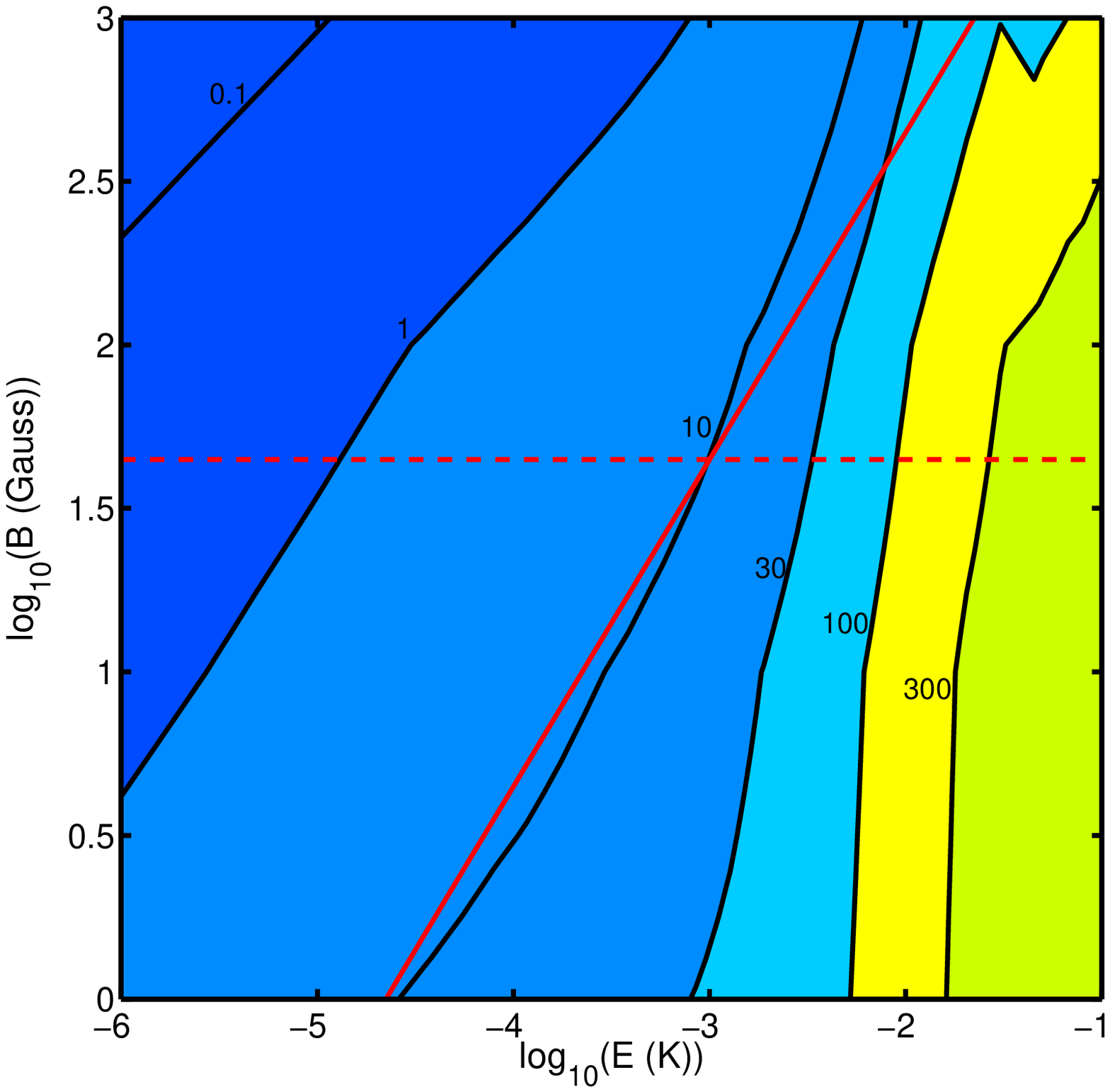}}
\\
\subfigure{\includegraphics[width=0.3\textwidth]{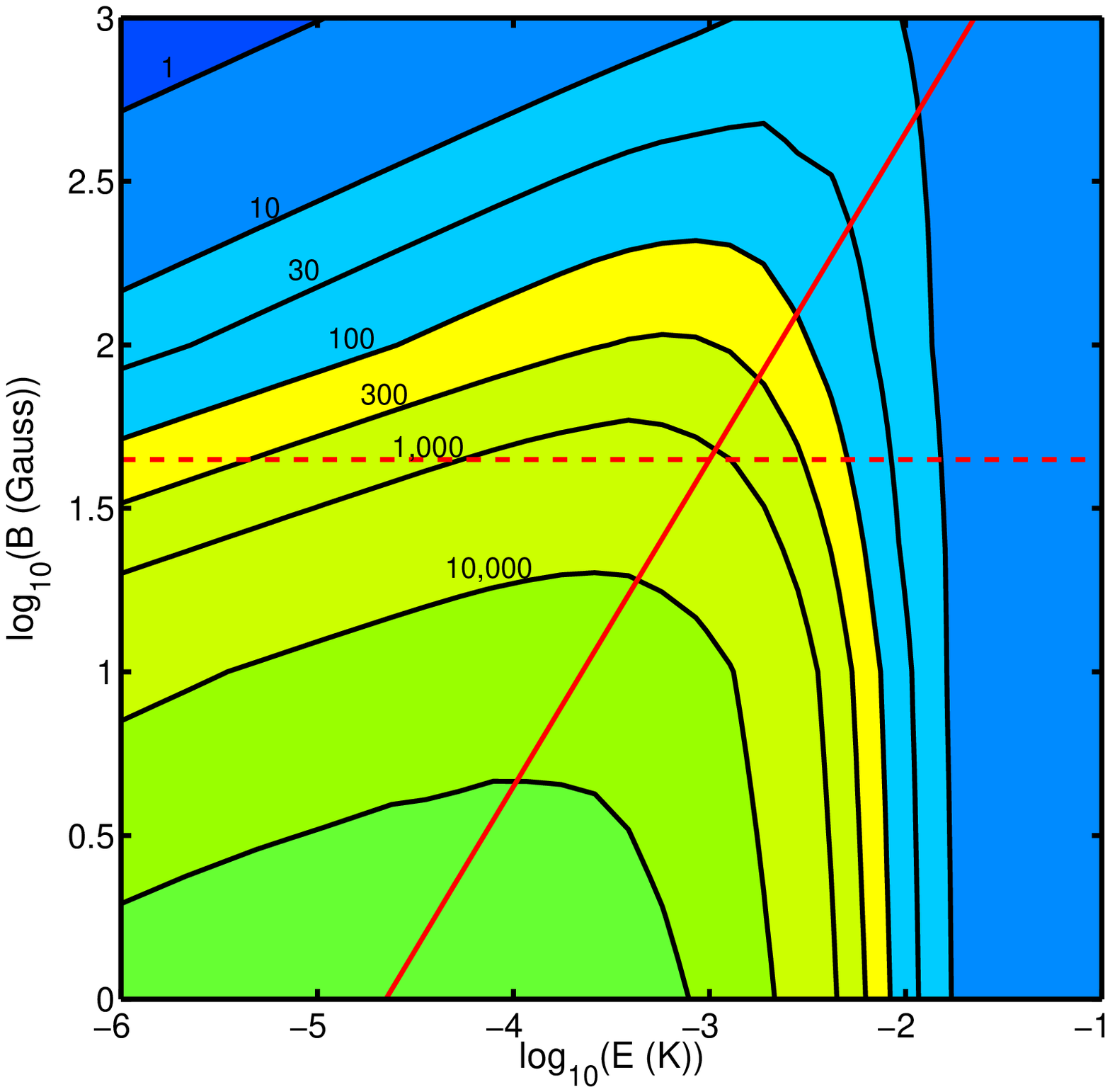}}
\qquad
\subfigure{\includegraphics[width=0.3\textwidth]{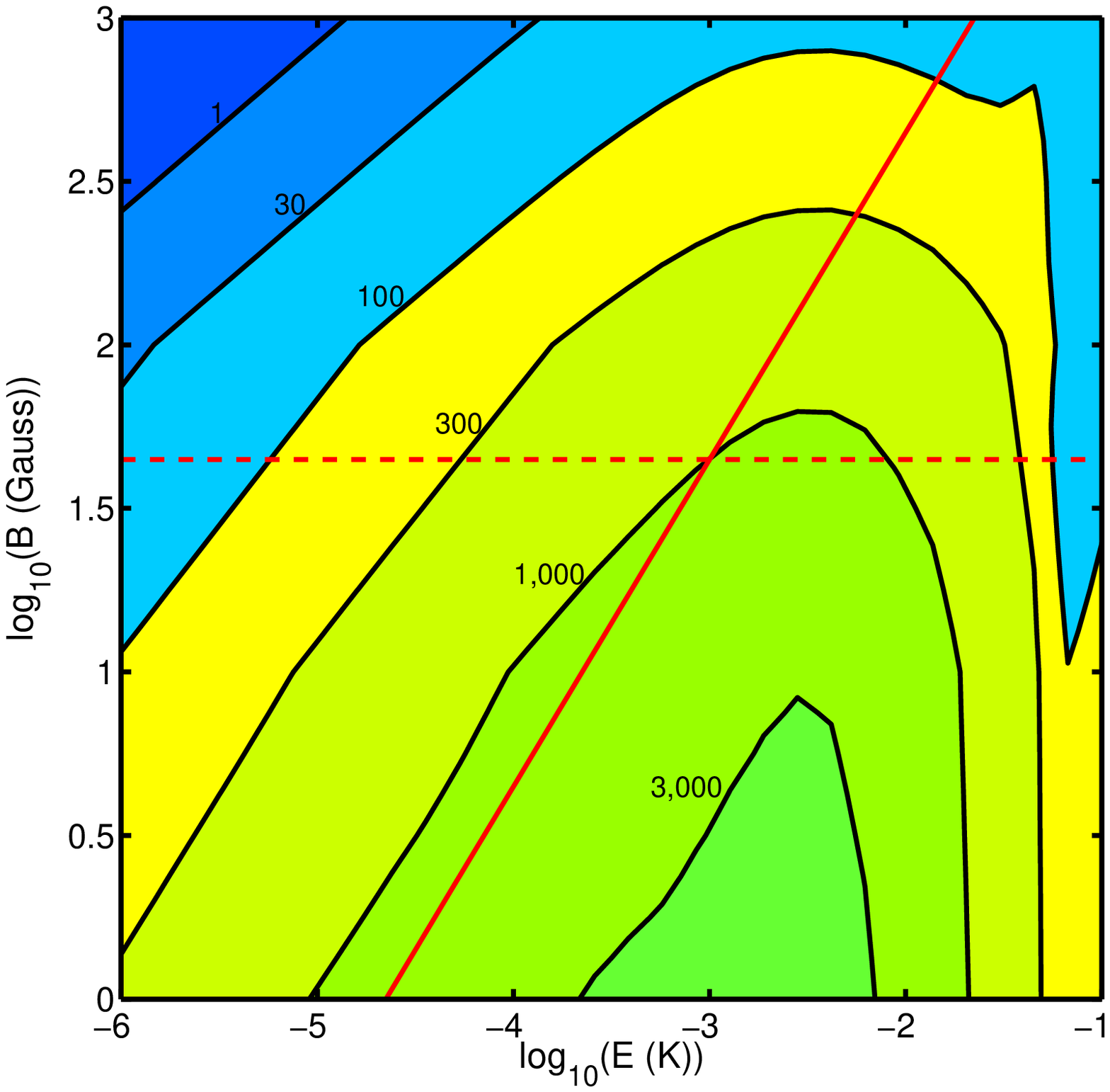}}
\qquad
\subfigure{\includegraphics[width=0.3\textwidth]{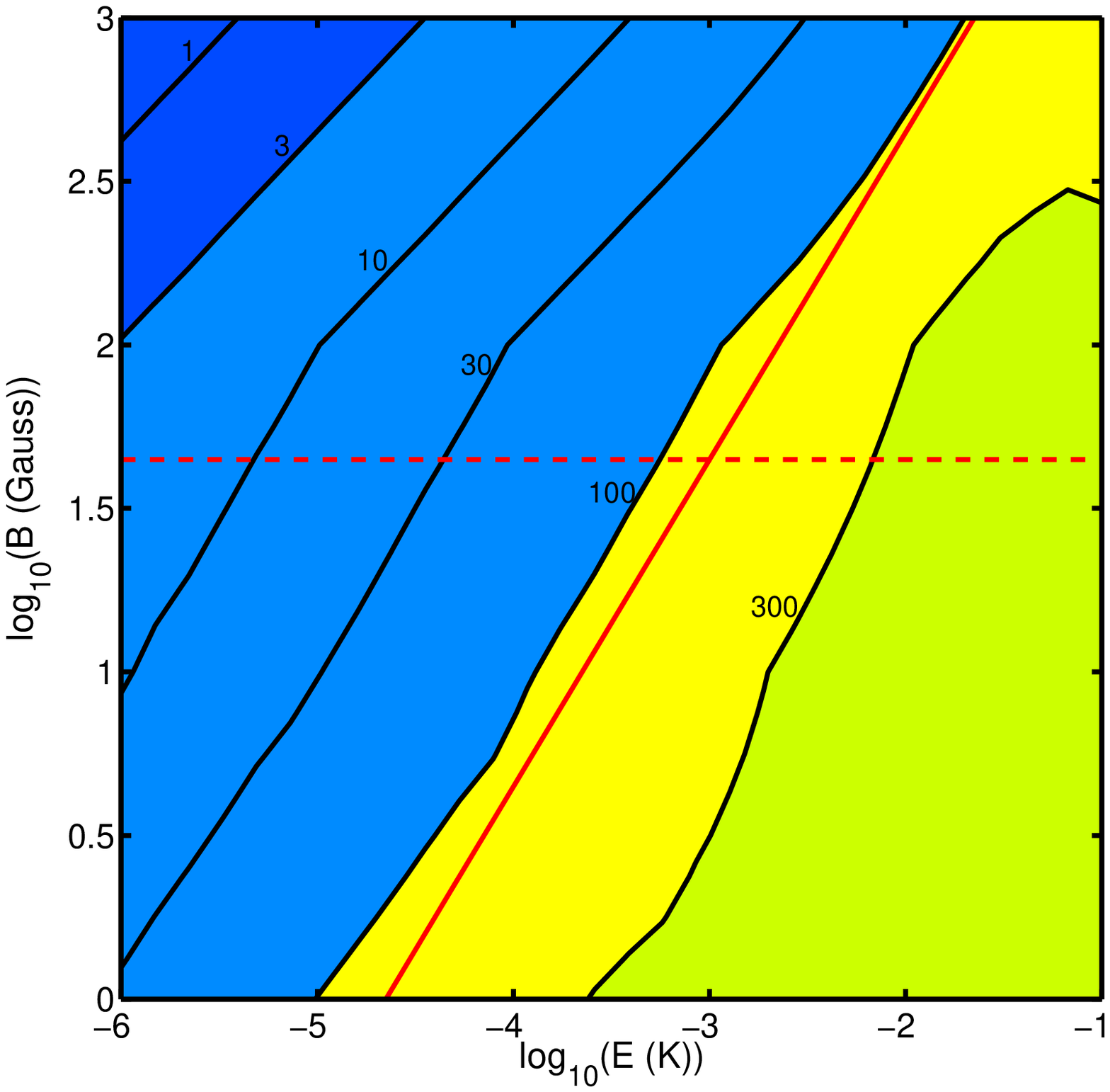}}
\caption{(color online). Contour plots of the ratio $\gamma$ of
the elastic to total inelastic cross sections including all s,
p and d-wave contributions. Contour plots are shown for six
different values of the potential scaling factor (clockwise
from the top left): (a) $\lambda_{\rm scl} =1$,
$a=-12.5\bar{a}$. (b) $\lambda_{\rm scl} =0.9915$,
$a=-\bar{a}$, (c) $\lambda_{\rm scl} =0.97$, $a=0.1\bar{a}$,
(d) $\lambda_{\rm scl} =0.96$, $a=0.3\bar{a}$, (e)
$\lambda_{\rm scl} =0.915$, $a=\bar{a}$, and (f) $\lambda_{\rm
scl} =1.00078$, $a=12.5\bar{a}$, The solid red lines show the
maximum field sampled by the spin-stretched state of the
molecules, $B=6k_{\rm B}T_{\rm NH}/g_e\mu_{\rm B}$ and the
dashed red line shows a similar boundary for Li atoms at
$T_{\rm Li}=1$~mK.} \label{fig:conts}
\end{center}
\end{figure*}

In a simple hard-sphere model of sympathetic cooling,
neglecting inelastic collisions, the temperature relaxes
towards equilibrium and reaches a 1/$e$ point after
$(m_1+m_2)^2/2m_1m_2$ collisions \cite{deCarvalho:1999}, where
$m_1$ and $m_2$ are the masses of the two species. For
sympathetic cooling to be successful we need the ratio of
elastic to inelastic cross sections to be much larger than
this. The general rule of thumb is that the ratio of the
elastic to the total inelastic cross sections, $\gamma=
\sigma_{\rm elas}/\sigma_{\rm inelas}$, must be greater than
about 100.

Figure~\ref{fig:L1Edep} shows the spin-stretched elastic and
total inelastic cross sections, incorporating all s, p, and d
partial wave contributions ($L=0,1,2$) for the original {\em ab
initio} potentials with $\lambda_{\rm scl}=1$ and for
representative scaled potentials with $\lambda_{\rm scl}=0.96$
and 0.915, chosen to give $a=0.3\overline{a}$ and
$a=\overline{a}$, respectively. The potential with
$\lambda_{\rm scl}=1$ has $a=-12.5\bar{a}$, and thus an
atypically large value of the elastic cross section. The spd
basis set gives convergence of the partial-wave sum in cross
sections for collision energies up to about 60~mK. At low
magnetic fields the kinetic energy release (i.e.\ the Zeeman
splitting between the incoming and outgoings channels) is less
than the height of the barriers in the outgoing channels, so
that the inelastic cross section is suppressed. As the magnetic
field is decreased, the Zeeman splitting decreases, the
centrifugal suppression increases, and the inelastic cross
section decreases. The dominant elastic processes are $\Delta
L=0$, so the elastic cross section does not display significant
field-dependence. It may also be seen from
Figure~\ref{fig:L1Edep} that the ratio $\gamma$ is in excess of
100 for low magnetic fields and a wide range of collision
energies up to about 30 mK. However, due to the presence of the
intermolecular spin-spin interaction
\cite{Janssen:NHNH:field:2011}, the reduction of the inelastic
cross section at low fields is not as dramatic as in Mg+NH
\cite{Wallis:PRL:MgNH:2009}.

To assess the prospect of sympathetic cooling over the range of
possible potentials, Figure~\ref{fig:conts} shows contour plots
of $\gamma$ as a function of the collision energy and magnetic
field strength for a variety of $\lambda$ values, chosen to
give a range of scattering lengths from $-12.5\bar{a}$ to
$+12.5\bar{a}$. It may be seen that the majority of possible
potentials give favourable values of $\gamma$ over a wide range
of fields and energies. The unfavourable potential at the upper
right is also atypical: it has $a=0.1\bar{a}$, so that the
s-wave elastic scattering cross section is a factor of 100
below the typical value of $4\pi\bar{a}^2$. Such unfavourable
behaviour might occur for any system, but is not very likely.

In an unbiased magnetic quadrupole trap, with a field zero at
the center, the molecules in the $m_{s_{\rm NH}}=1$ state at a
temperature $T_{\rm NH}$ will have a Boltzmann density
distribution
\begin{equation}
\frac{\rho}{\rho_0}=\exp\left(-\frac{g_e\mu_{\rm B} Bm_{s_{\rm NH}}}{k_{\rm B}T_{\rm NH}}\right).
\end{equation}
At a given temperature on the energy axis of the contour plots
in Figure~\ref{fig:conts} less than 1\% of molecules will
experience a magnetic field greater than $B=6k_{\rm B}T_{\rm
NH}/g_e\mu_{\rm B}$, which is shown as a solid red line. If the
atomic cloud occupied the same volume as the molecular cloud,
the large majority of collisions would occur at energy/field
combinations below this line. However, at the start of
sympathetic cooling the Li is likely to be substantially colder
than the NH, probably 1~mK or less. The Doppler limit for Li is
140~$\mu$K. The magnetic moment of $^7$Li in its $F=2$, $M_F=2$
state is only half that of NH, but nevertheless it is likely
that the Li cloud will be significantly smaller. Li-NH
collisions will take place only where both species are present,
near the centre of the trap where fields are low. This will
reduce the frequency of collisions but may nevertheless be
advantageous because it will mean that collisions mostly take
place in regions well below the solid red line. Over 99\% of Li
atoms at a temperature $T_{\rm Li}$ will occupy regions of the
trap with fields less than $B=12k_{\rm B}T_{\rm Li}/g_e\mu_{\rm
B}$, and this boundary is shown in Figure ~\ref{fig:conts} as a
dashed red line for Li at 1~mK. For many potentials the
inelasticity in this region remains low to temperatures above
100~mK at fields below 100~G.

It can be seen from Figure~\ref{fig:conts} that the contour
plots depend mostly on the magnitude of the scattering length
and not on its sign, because the s-wave elastic cross section
is proportional to the square of the scattering length. There
are small deviations from this: for $a\approx\bar{a}$ there is
a p-wave resonance around 50~mK, and for $a\approx-\bar{a}$
there is a d-wave resonance around 50 mK and 500 G. If the real
potential is such that $a$ is accidentally close to zero, such
as for $\lambda_{\rm scl}=0.97$, sympathetic cooling will not
succeed because of the small elastic cross section. However as
$|a|/\bar{a}$ increases to ``typical" values close to 1 (such
as for $\lambda_{\rm scl}=0.96$), sympathetic cooling has a
good chance of success. For all potentials with
$|a|/\bar{a}\ge1$ it can be seen that, once sympathetic cooling
has started and the sample of molecules is cooled to
sub-millikelvin temperatures, $\gamma$ increases and the
molecules become increasingly stable to spin-relaxation. The
temperature at which sympathetic cooling begins to be effective
depends quite strongly on the potential, but for some
potentials it is as high as 100~mK.

Changing the reduced mass of the collision system has an effect
quite similar to scaling the potential
\cite{Zuchowski:NNH:2011}. Thus if the scattering length for
one isotopic combination proves to be too small to allow
sympathetic cooling to succeed, another combination may well be
more successful.

A further important consideration in a sympathetic cooling
experiment concerns the absolute collision rate. As seen above,
the elastic cross section in the ultracold regime fluctuates
dramatically near Feshbach resonances but its ``background"
value is $4\pi\bar{a}^2$, which for Li+NH is 1365~\AA$^2$. The
elastic cross section drops off at higher energy for some
potentials, but in most cases it does not change by more than a
factor of 3 from $4\pi\bar{a}^2$ (in either direction) at
collision energies around 100~mK. At this temperature, with a
reasonable Li atom density of $10^{10}$ cm$^{-3}$ and an
elastic cross section of $10^3$~\AA, each molecule will
experience a collision with a Li atom once every 0.8~ms.
However, this time will be extended if only part of the
molecular cloud overlaps with the atomic trap. In a magnetic
quadrupole trap with a molecular temperature much higher than
the atomic temperature, the overlap between the two clouds will
be quite poor. This might be alleviated by using a trap of
Ioffe-Pritchard type, with a finite magnetic field at the
bottom of the trap. This would have the beneficial side-effect
of preventing Majorana losses \cite{Majorana:1932,Petrich:1995}
at the center of the trap. As cooling progresses, the molecular
cloud will collapse towards the center of the trap, so the
overlap with the atomic trap will improve. As this occurs, the
bias field of the Ioffe-Pritchard trap can be reduced to
increase the density and the corresponding collision rate.

\section{Conclusions}

We have calculated the quartet potential energy surface for NH
interacting with Li and used it to calculate elastic and
inelastic cross sections for collisions between cold NH and Li
in magnetically trappable spin-stretched states. The potential
energy surface is highly anisotropic, with a well depth about
1800 cm$^{-1}$ at the Li-NH geometry but only 113 cm$^{-1}$ and
the NH-Li geometry. The strong anisotropy drives
spin-relaxation collisions that are considerably faster than in
systems such as Mg-NH and He-NH, but are nevertheless
suppressed by centrifugal barriers when both species are in
spin-stretched states.

We have explored the dependence of the collision properties on
the potential energy surface, or equivalently on the reduced
mass of the colliding system. In the ultracold regime, both the
elastic and inelastic (spin-relaxation) cross sections show
dramatic fluctuations as the potential is varied, due to
Feshbach resonances that occur as a bound state passes through
zero energy. The potential-dependence is considerably reduced
at higher energies, when higher partial waves contribute to the
scattering. The major effect of using an unconverged basis set
in the scattering calculations is to shift the positions of the
resonances without changing their general behaviour.

We have calculated the ratio of elastic and inelastic
(spin-relaxation) cross sections, as a function of collision
energy and magnetic field, for a variety of potential energy
surfaces. This ratio is key to the success of sympathetic
cooling, because spin-relaxation cross sections convert
internal energy into relative translation of the colliding
species and thus cause trap loss. We find that most (but not
all) potential energy surfaces produce ratios that are
favorable for sympathetic cooling, provided the molecules can
be precooled to a temperature around 20~mK. The exceptions are
potentials where the scattering length is accidentally near
zero, which produce small elastic cross sections in the
ultracold regime. If this unlucky situation occurs in practice,
it can probably be avoided by using a different isotopic
combination of NH and/or Li.

It is important to consider the overlap between the atomic and
molecular traps. If both species are held in a magnetic
quadrupole trap and the atoms are initially much colder than
the molecules, the atomic cloud will be much smaller than the
molecular cloud. This may be advantageous because it means that
atom-molecule collisions will take place only at the low
magnetic fields sampled by the atoms, and for many potentials
such low-field collisions have low inelasticity even at
temperatures around 10~mK. However, there is a price to pay
because poor overlap extends the time needed for sympathetic
cooling. This might be alleviated by using a Ioffe-Pritchard
trap to increase the size of the atomic cloud in the early
stages of cooling.

\section*{Acknowledgments}

We are grateful to Liesbeth Janssen for valuable discussions.
This work is supported by EPSRC under the collaborative project
CoPoMol of the ESF EUROCORES Programme EuroQUAM.

\bibliographystyle{epj}
\bibliography{../../all,../conical-photon/thref}

\end{document}